\documentclass[%
 reprint,
amsmath,
amssymb,
aps,
pra,
]{revtex4-2}

\usepackage{graphicx}
\usepackage{dcolumn}
\usepackage{bm}
\usepackage{siunitx}
\usepackage[T1]{fontenc}
\usepackage[mathlines]{lineno}
\usepackage[version=4]{mhchem}
\usepackage{booktabs}

\begin{document}

\preprint{APS/123-QED}

\title{The carbon cost of materials discovery: Can machine learning really accelerate the discovery of new photovoltaics?}

\author{Matthew Walker}%
\email{matthew.walker.21@ucl.ac.uk}
\affiliation{%
Department of Chemistry, University College London, 20 Gordon Street, London WC1H 0AJ,
United Kingdom
}%

\author{Keith T.  Butler}
\email{k.t.butler@ucl.ac.uk}
\affiliation{%
Department of Chemistry, University College London, 20 Gordon Street, London WC1H 0AJ,
United Kingdom
}

\date{\today}

\begin{abstract}
Computational screening has become a powerful complement to experimental efforts in the discovery of high-performance photovoltaic (PV) materials. Most workflows rely on density functional theory (DFT) to estimate electronic and optical properties relevant to solar energy conversion. Although more efficient than laboratory-based methods, DFT calculations still entail substantial computational and environmental costs. Machine learning (ML) models have recently gained attention as surrogates for DFT, offering drastic reductions in resource use with competitive predictive performance. In this study, we reproduce a canonical DFT-based workflow to estimate the maximum efficiency limit and progressively replace its components with ML surrogates. By quantifying the \ce{CO_2} emissions associated with each computational strategy, we evaluate the trade-offs between predictive efficacy and environmental cost. Our results reveal multiple hybrid ML/DFT strategies that optimize different points along the accuracy–emissions front. We find that direct prediction of scalar quantities, such as maximum efficiency, is significantly more tractable than using predicted absorption spectra as an intermediate step. Interestingly, ML models trained on DFT data can outperform DFT workflows using alternative exchange–correlation functionals in screening applications, highlighting the consistency and utility of data-driven approaches. We also assess strategies to improve ML-driven screening through expanded datasets and improved model architectures tailored to PV-relevant features. This work provides a quantitative framework for building low-emission, high-throughput discovery pipelines.
\end{abstract}

\maketitle


\section{\label{sec:level1} INTRODUCTION}

The development of new material functionalities has historically been a time-consuming process, with new materials or materials' applications often discovered serendipitously or taking many decades of careful synthesis and characterization before realizing a real-world application~\cite{cheetham_chemical_2022}. In recent years, with the increase in computing power and the sophistication of materials modelling software, computational chemistry has promised to accelerate this process by providing qualitative insight and design rules as well as quantitative predictions allowing virtual screening of new materials for given applications. However, atomistic modelling has known limitations for the discovery of new materials.  In recent years, the emergence of data-driven approaches, notably machine learning (ML), and the availability of large, high-quality annotated datasets of material properties have been predicted to be a route to accelerate computational materials design. But there are many open questions for applying ML to PV discovery: Are these methods really reliable? Which methods are best suited to the task? How does the quality of the underlying data affect model performance? Finally, what should we actually model? In this paper, we set out to address some of these questions in the context of photovoltaic (PV) materials discovery. and provide some concrete guidelines on how effective ML is for PV discovery currently by estimating the carbon cost of both ML-based and density functional theory (DFT)-based materials discovery. We also provide suggestions on how to advance in the future.

Global PV capacity reached approximately 1.6 TW in 2023~\cite{massongaetan_snapshot_2024}, and a future push toward 30--70 TW by 2050 could see PVs meeting most of the world's energy requirements~\cite{haegel_terawatt-scale_2019}. Achieving this target requires the development of new materials as well as the optimization of existing ones~\cite{blakesley_roadmap_2024}. While crystalline and multi-crystalline Si modules remain the industrial standard~\cite{battaglia_high-efficiency_2016}, alternative materials such as amorphous Si~\cite{sayed_optimization_2025}, CIGS~\cite{ramanujam_copper_2017}, CdTe~\cite{scarpulla_cdte-based_2023}, organic photovoltaics~\cite{solak_advances_2023}, and dye-sensitized solar cells~\cite{oregan_low-cost_1991} have been commercialized to varying degrees of success. A number of perovskites have also emerged as promising candidates in the last decade~\cite{green_emergence_2014}. However, established technologies often rely on critical raw materials, toxic elements, or suffer from long-term stability issues, conversion efficiency limitations, or low technological flexibility; overcoming these challenges is essential for reaching TW-level production of PV energy~\cite{zakutayev_emerging_2021}.

New inorganic materials offer significant promise as future PV absorbers due to their potential for low-cost fabrication, defect tolerance, earth abundance, and facile synthesis via various techniques such as sol-gel processing or sputtering~\cite{needleman_economically_2016, yang_above-bandgap_2010, wu_narrow_2022,welch_trade-offs_2017}. These materials exhibit stability across a wide range of thermal, chemical, and mechanical conditions and are compatible with device architectures that may offer lower capital costs, enabling rapid scale-up~\cite{needleman_economically_2016}.

Computational modelling has played an important role in the development of new inorganic photovoltaic materials such as CZTS~\cite{yadav_simulation_2025,ahmad_computational_2022}, SnS~\cite{vidal_band-structure_2012}, BiSI~\cite{mganose_relativistic_2016}, Sb$_2$Se$_3$~\cite{wang_upper_2024}, CdTe~\cite{yang_review_2016} and many others. Typically, these studies have been DFT calculations allowing accurate estimation of optical absorption, carrier transport and defect properties~\cite{yuan_discovery_2024}. Although these DFT calculations are more efficient than experimental synthesis and characterisaton, they nonetheless have a non-negligible energy cost. In recent years there has been a trend to replace some of the costly DFT calculations with ML surrogate models. However, the questions previously raised about the veracity of these models remain largely unanswered.

To address these questions, we have developed a framework that enables the joint assessment of both predictive accuracy and carbon emissions associated with different computational approaches for estimating PV performance in novel inorganic crystalline materials. These approaches span from hybrid-functional DFT (the most computationally expensive) to direct ML estimation of maximum PV efficiency (the least expensive), and include intermediate strategies such as predicting optical absorption profiles or applying corrections to low-fidelity DFT calculations based on the generalized gradient approximation (GGA)~\cite{perdew_generalized_1996}. The paper begins with a detailed outline of our evaluation methodology, covering both PV property estimation and carbon emission quantification. We then compare these approaches in terms of predictive efficacy and environmental cost. Our analysis allows us to propose optimal trade-offs, highlight important limitations, and suggest promising directions for future research aimed at improving the effectiveness and sustainability of computational PV screening. More broadly, our framework offers a template for evaluating computational discovery pipelines in which resource intensity is considered alongside predictive performance—a consideration we believe will be increasingly important across many areas of energy research.

\section{Evaluation methodology}

We provide details of the different design choices in our evaluation protocol. Covering approaches to obtain the carbon emissions of calculations, the optical absorption spectra and the maximum PV efficiency.

\subsection{The Carbon Cost of Discovery}

Ultimately, we are interested in developing new photovoltaic materials as a renewable energy technology. Therefore, it is important to consider the energy cost of a discovery campaign. While computational discovery is less resource intensive than experimental programmes, it is not carbon neutral. One promise of ML is that it can reduce the computational cost and ultimately the resource required for discovery. To assess the trade-off between prediction accuracy and carbon cost of the calculations involved, we have used the \verb|CodeCarbon| package~\cite{courty_mlco2codecarbon_2024}, which integrates into computational workflows to estimate the \ce{CO_2} emissions associated with running a given job by monitoring total energy usage across all processing units. This enables facile comparison of computational chemistry calculations, typically CPU-based, and ML inferences, which mostly use GPUs. \ce{CO_2} emissions are then estimated based on the sources of energy for the grid in the location of the computer, in our case the UK.  The specific numbers for these emissions are thus very sensitive to change, so we mostly give relative emissions. However, we provide some raw numbers to give context for the scale of the \ce{CO_2} emissions associated with the calculations in this work.

\subsection{Calculating Maximum Efficiencies}

The computational design of materials typically relies on the availability of a readily computable figure of merit (FoM), which provides a measure of how good a given material is for an application. In photovoltaics, the detailed balance limit~\cite{shockley_detailed_1961} gives the maximum achievable power conversion efficiency of a single-junction PV cell as a function of the band gap. This simple FoM assumes a step-function absorptance $\left(A(E)\right)$, which is particularly inaccurate for indirect band-gap absorbers, which typically show a more gradual onset of absorption.

Instead, the efficiency of potential PV absorber materials can be estimated using the spectroscopic limited maximum efficiency(SLME)~\cite{yu_identification_2012}. The theory and practical details of calculating SLMEs are discussed in the following section. To distinguish between these methods (since both use detailed balance), we shall henceforth refer to methods using a step-function approximation of the absorption spectrum as `step-function methods' and those that use the calculated/predicted spectrum as `SLME methods'.

The SLME of a material requires an absorption profile $\alpha(E)$, usually in units of \unit{\per\centi\metre}, and an `offset' (our taxonomy):
\[\Delta = E_g^{da}-E_g,\]
where $E_g^{da}$ is the minimum direct, dipole-allowed band gap and $E_g$ is the fundamental band gap which, in contrast, may be indirect and dipole-forbidden. 
The \textit{absorptance}, $A(E)$, for a material of thickness $d$ is calculated from the \textit{absorption}, $\alpha(E)$, using a Lambert-Beer approximation:
$$A(E)=1-\exp\left(-2d\cdot\alpha\left(E\right)\right).$$
This is used to calculate the short-circuit current (density):
\[J_\text{SC}=e\int_0^\infty{A(E)\phi_\text{AM1.5G}(E)dE},\]
where $\phi_\text{AM1.5G}(E)$ is the spectrum of solar radiation received at ground level on Earth when the Sun is perpendicular and $e$ is the charge on an electron. The internal quantum efficiency is assumed to be one: that is, all photons absorbed contribute to the current.

Detailed balance says that the rate of absorption photon emission from radiative recombination processes must equal the photon absorption from the surroundings, which can be quantified using the black-body spectrum at the temperature, $T$, of the solar device. This gives the reverse saturation current density (or recombination current density) as
\[J_0^{\text{rad}}=\frac{e\pi}{f_r}\int_0^\infty \:A(E)\phi_\text{BB}(E)dE,\]
with the black-body spectrum given by
\[\phi_\text{BB}(E)=\frac{2e\pi}{h^3c^2}\int_0^\infty \frac{A(E)E^2}{\exp\frac{E}{k_BT}-1}dE,\]
and
\[f_r=\exp\left(\frac{-\Delta}{k_BT}\right),\]
which uses the offset defined above and represents the fraction of recombination due to radiative processes.

The voltage-dependent total current density is then multiplied by the voltage to give the power: 
\[P=VJ=V\left\{J_\text{{SC}}-J_0^{\text{rad}}\left[\exp\left(\frac{eV}{k_BT}\right)-1\right]\right\}.\]
The maximum value of this power, $P_\text{max}$ will be found at some balance of $V$ and $J$, giving the optimal efficiency as \[\eta=\frac{P_\text{max}}{P_\text{in}}\] where the denominator is given by integrating over the AM1.5G spectrum, which has by convention been normalised to integrate to approximately 1000 \unit{\watt\per\metre\squared}. 

\subsection{Optical absorption calculations}
The SLME is calculated using the absorption profile of the material, which is directly accessed from electronic structure calculations, such as DFT. The reliability of the calculated SLME thus depends on the reliability of the underlying absorption profile. In general, more accurate DFT (or other electronic structure methods) calculations of the absorption profile require more computational resource to calculate. A simple hierarchy of electronic structure methods could include, in increasing accuracy and cost, the generalised gradient approximation~\cite{perdew_generalized_1996} (GGA) to DFT, hybrid methods such as the Heyd–Scuseria–Ernzerhof~\cite{heyd_hybrid_2003} (HSE) functional, and GW routines. However, intermediate methods exist: particularly relevant to this study is the process of applying a scissor correction to a low-fidelity (e.g. GGA) spectrum using the difference in band gaps calculated at the low-fidelity and a higher-fidelity (e.g. HSE) level. We can consider applying a GGA$\to$HSE scissor correction, 
\[\Delta E=E_g^\text{HSE}-E_g^\text{GGA},\] 
to a GGA absorption spectrum as an approximation to an HSE-level absorption spectrum:
\[\alpha_\text{HSE}\approx\alpha_\text{GGA}(E-\Delta E),\]
which has the effect of shifting the spectrum to the right in most cases, since hybrid band gaps are usually larger than their GGA equivalents.
This approach to approximating HSE spectra has been shown to be reasonable by Yang et al.~\cite{yang_high-throughput_2022}, who also showed that the independent particle approximation (IPA) to optics calculations produces spectra that generally agree well with those calculated using the more rigorous (and expensive) random-phase approximation (RPA). 
The fundamental band gaps needed for the scissor correction and offset are typically calculated using band structure calculations, while optics calculations provide both the dielectric tensor and transition dipole matrix required for $\alpha(E)$ and $E_g^{da}$, respectively.
\section{RESULTS}
In order to accelerate the identification of new materials with promising SLME values, one can propose replacing computationally demanding electronic structure calculations with surrogate models trained on existing data and capable of making predictions at a fraction of the cost, indeed this is done quite routinely~\cite{de_angelis_impact_2023,witman_defect_2023,woods-robinson_designing_2023}. We have trained surrogate models for each of the electronic structure steps in the SLME calculation workflow and now assess (i) how accurate these models are and (ii) how the errors in the model predictions propagate and affect the final ranking of new materials in the calculated SLME and (iii) the relative carbon cost of the different approaches. 

In Table \ref{methods_table} we provide a list of potential workflows where electronic structure calculations are replaced with ML surrogates. We have also provided a number of workflows that include approaches using the detailed balance limit and purely GGA-level properties to provide context for the accuracies and costs of the machine-learning-based approaches. Note that in Method II the model has been trained on scissor-corrected spectra, so a subsequent calculated or predicted scissor correction is not necessary. Those properties that are calculated using straightforward and negligibly expensive operations, in our case executed in Python, are represented as `Py' in the table. For instance, detailed balance approaches use Python to estimate an absorptance profile from the material's band gap. 

Strictly speaking, the DFT-based $\alpha(E)$ is also calculated in Python from the dielectric data in the output files of a computational chemistry calculation, so this column is a stand-in for a calculation or prediction that produces an absorption spectrum. We have also chosen to use GGA-level band gaps for the offset calculation: this is because the offset requires an optics calculation, so if the offset is calculated then you would also calculated a GGA absorption spectrum in the process, so it wouldn't be worth predicting one and not the other. GGA offsets will introduce some error, though both gaps in the equation will be wrong by similar amounts, cancelling out some of this error. However, the test dataset used GGA-level offsets, so this source of error was not examined in this work. 

\begin{table}[!ht]
    \centering
    \begin{tabular}{l l l l l l l l}
    \toprule
        \textbf{Method} & \textbf{$E_G$} & $E_G^\text{DA}$ & $\alpha(E)$ & $A(E)$ & $\Delta E$ & $\Delta$ & SLME \\ \midrule
        I & - & - & - & - & - & - & ML \\ 
        II & - & - & ML & Py & - & ML & Py \\ 
        III & GGA & GGA & GGA & Py & ML & Py & Py\\ 
        IV & ML & - & - & Py & - & - & Py \\ 
        V & HSE & - & - & Py & - & - & Py \\ 
        VI & GGA & - & - & Py & ML & - & Py \\ 
        VII & GGA & GGA & GGA & Py & - & Py & Py \\ 
        VIII & HSE & GGA & GGA & Py & Py & Py & Py \\ \bottomrule
    \end{tabular}
    \caption[Table 1]{Table outlining the methods considered for estimating SLMEs, either directly or via the properties required to calculate SLMEs, with properties originating from DFT calculations at GGA and HSE level or ML predictions.}
    \label{methods_table}
\end{table}

\begin{figure}[htbp!]
\centering
\scalebox{0.5}[0.5]{
\includegraphics[width=0.9\textwidth]{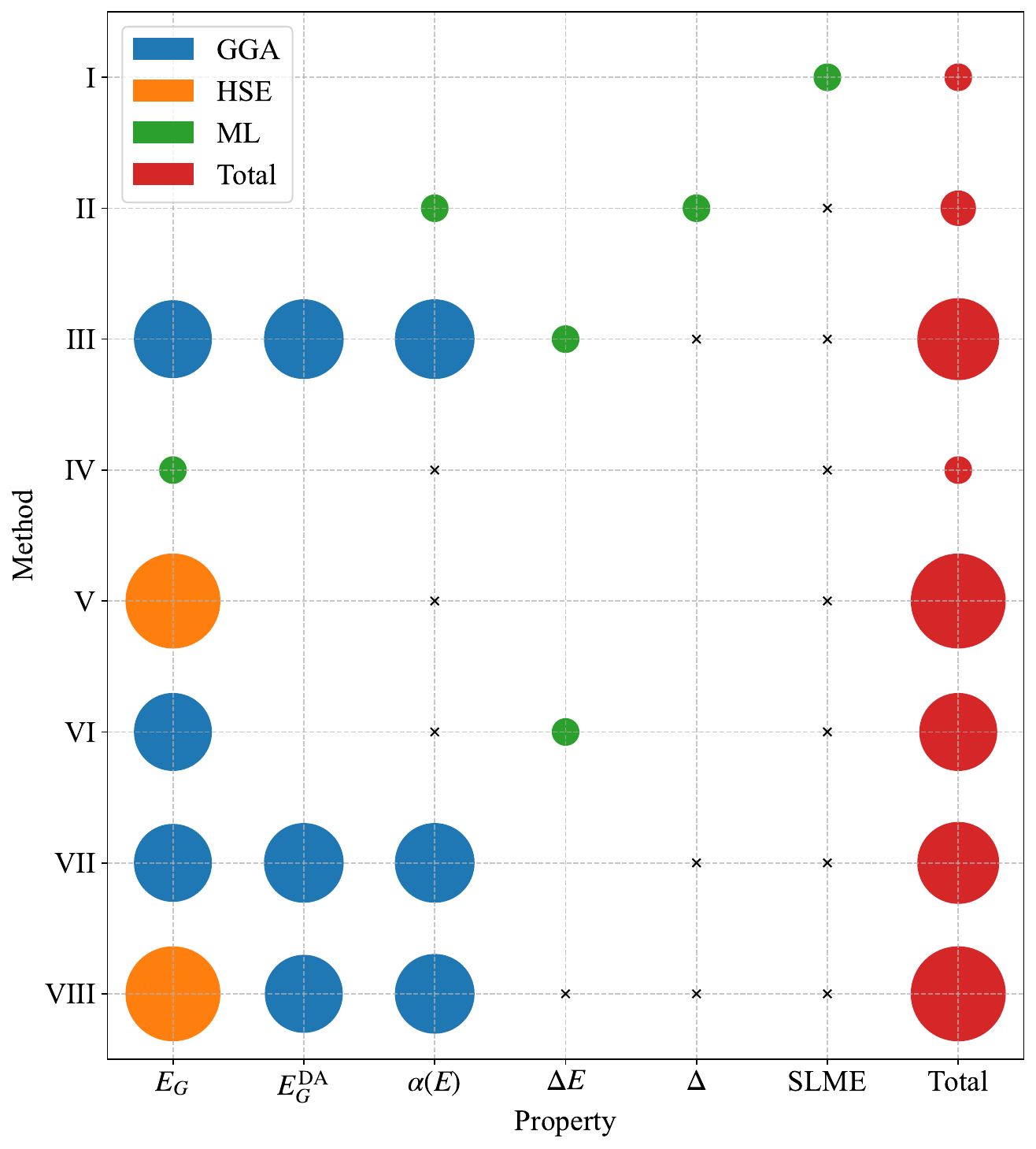}
}
\caption[Fig5]{Plot of the methods for estimating SLMEs outlined in Table \ref{methods_table}, with crosses representing Python calculations and circles representing more costly calculations, with area $A\propto \ln \left(\frac{C}{C_\text{min}}\right)$ for cost $C$. The absorptance column from Table \ref{methods_table}} has been excluded for brevity.
\label{methods_costs}
\end{figure}

Figure \ref{methods_costs} shows the relative cost of the calculations and predictions used in this work. Note that the area of a circle is proportional to the natural logarithm of its relative carbon cost, so the difference is even more stark than it appears. The negligible Python calculations are given as crosses to emphasise their low cost. ML inferences are also extremely inexpensive, though can be more meaningfully quantified as incurring a carbon cost around 1/2000x that of a static GGA calculation, which is itself around an order of magnitude cheaper than a similar HSE calculation. In terms of energy, this single ML inference used around $1.9\times 10^-3$ Wh (around 7 J), which \verb|CodeCarbon|~\cite{courty_mlco2codecarbon_2024} estimates as producing $4.5\times10^{-4}$ g of \ce{CO_2}: equivalent to driving a typical diesel transit bus 0.3 mm~\cite{hesterberg_comparison_2008}.

Finally, optics and band structure calculations are more expensive than static calculations, making an accurate absorption spectrum predicting model all the more promising. The figure does in some ways under-represent the cost of machine learning approaches, since training (and hyperparameter tuning, though this was not performed in this work) is not included. Training Model 1 on the 4.8k dataset for 300 epochs was equivalent to $1.7\times10^5$ times that of a single inference. This is more indicative of how small the inference costs are than how large the training costs are. Moreover, these are one-off costs that would become negligible when considering the application of these models on vast datasets, and would not be incurred by future users of these models. 
\begin{figure*}[htbp!]
\centering
\scalebox{1.3}[1.3]{
\includegraphics[width=0.75\textwidth]{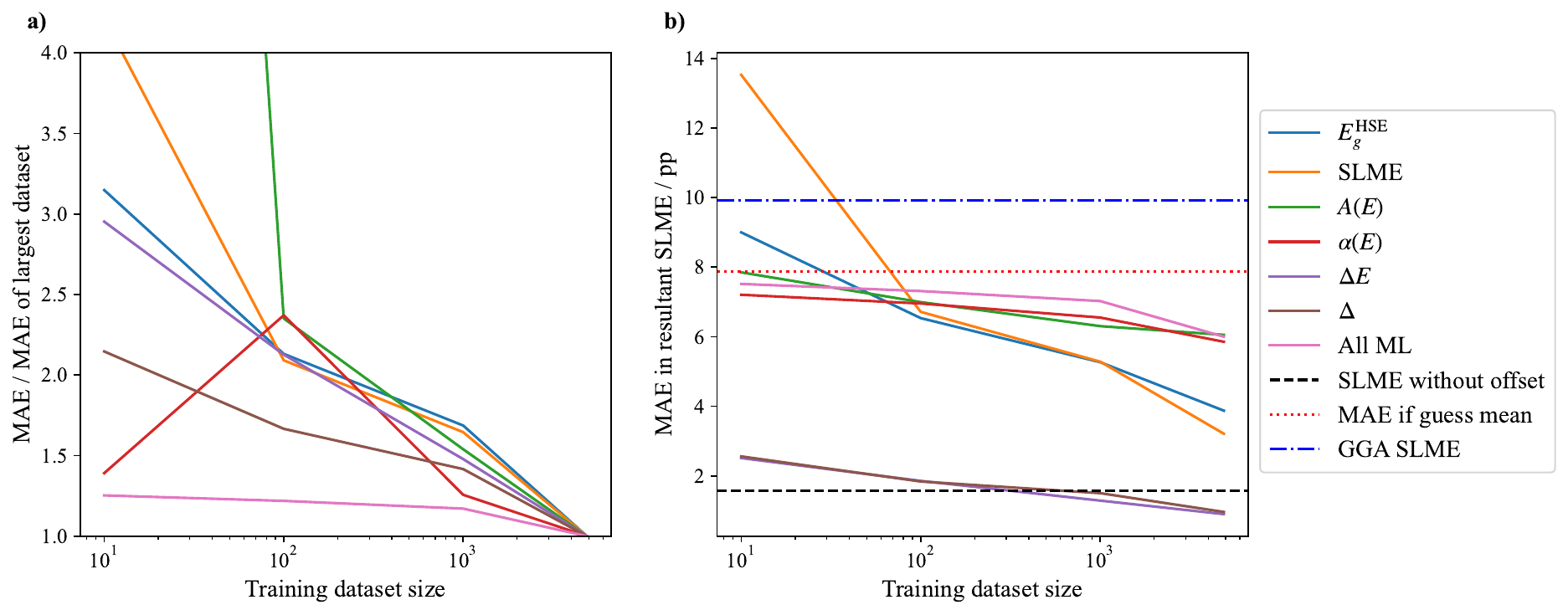}
}
\caption[Fig5]{Learning curves for each property, looking at a) relative errors for the given property and b) the resultant error when this learned property is used to calculate the SLME. For all but the full dataset, the values are averaged over three random sub-samples of the dataset. The errors are calculated using the test dataset used throughout the paper.}
\label{learning_curves}
\end{figure*}
\subsection{We haven't reached data saturation}
Before analysing the effects of different ML interventions in the estimation of SLMEs, we first look at how accurate the various ML models are and how their performance scales with training data. It is a well-known phenomenon that the performance of deep learning models generally scales very well with data~\cite{banko_scaling_2001,amodei_deep_2015, hestness_deep_2017, sun_revisiting_2017, bailly_effects_2022} and therefore we investigate how the models we use scale with the available data.

Figure~\ref{learning_curves}(a) shows how the performance of the various ML models on a held-out test set evolves as the size of the training data increases. The dataset size is truncated at just under 5,000: the number of materials in both the band gap and absorption spectra datasets (after a test/training split), since both are required to calculate a scissor-corrected SLME. From this plot, it is quite clear that all of the property models are still improving with more training data, and we have not reached data saturation. In the SI we show how the predicted absorption spectrum of GaAs (not in the training set) improves with more training data: in particular, point-to-point correlation is achieved at around 1k training data points, with the curve becoming smooth.

In the context of the final target (accurate SLMEs for PV screening), we show the effect of dataset size in Figure~\ref{learning_curves} (b). Here, the abscissa is the error in the final estimated SLME when a particular ML model is used in the workflow. The dotted red line shows a null hypothesis, where our ``model'' simply predicts the mean SLME of the training data. Clearly, with a few training data ($\leq 100$) all models exceed this baseline, even the model that predicts the high-dimensional absorption spectrum. The plot also demonstrates how with $\sim$100 data points all workflows incorporating ML perform favourably when compared to calculating an SLME from a low-cost, low-fidelity DFT optical absorption profile obtained from a GGA calculation (without a scissor correction). 

Perhaps more important than the absolute values in Figure~\ref{learning_curves} (b) are the gradients. The gradients give us an indication of how the predictions may be improved with additional data collection. The gradient of the direct prediction of SLME (with no DFT intermediates) shows the steepest gradient and extrapolation at the current rate of model improvement with several tens of thousands of high-quality estimates of SLME a model with negligible errors is possible.  

If the absorption spectrum is known but the offset is not, Figure \ref{learning_curves} suggests that the inclusion of an ML-predicted offset is worthwhile (rather than a semi-SLME approach with $f_r=0$), provided that the training dataset size exceeds $\sim10^3$.

Predicting the absorption spectra and calculating the absorptance from them gives errors very similar to predicting the absorptance directly. This is perhaps surprising as absorptance spectra are naturally scaled to be between 0 and 1, and are relatively featureless (all more or less sigmoid-shaped), whereas absorption values may be anywhere between 0 and $10^7$, and the overall profile is generally more irregular. One possible explanation is that, when the absorptance is calculated from the predicted absorption spectrum, small discrepancies are smoothed out by the exponential function, thereby reducing the propagated error in the final SLME, whereas direct absorptance prediction has no such advantage. Given the better performance with the full training set, the methods that included spectral prediction predicted absorption rather than absorptance. 

\subsection{Some methods are more worth pursuing than others}
\begin{figure}[htbp!]
\centering
\scalebox{0.5}[0.5]{
\includegraphics[width=0.9\textwidth]{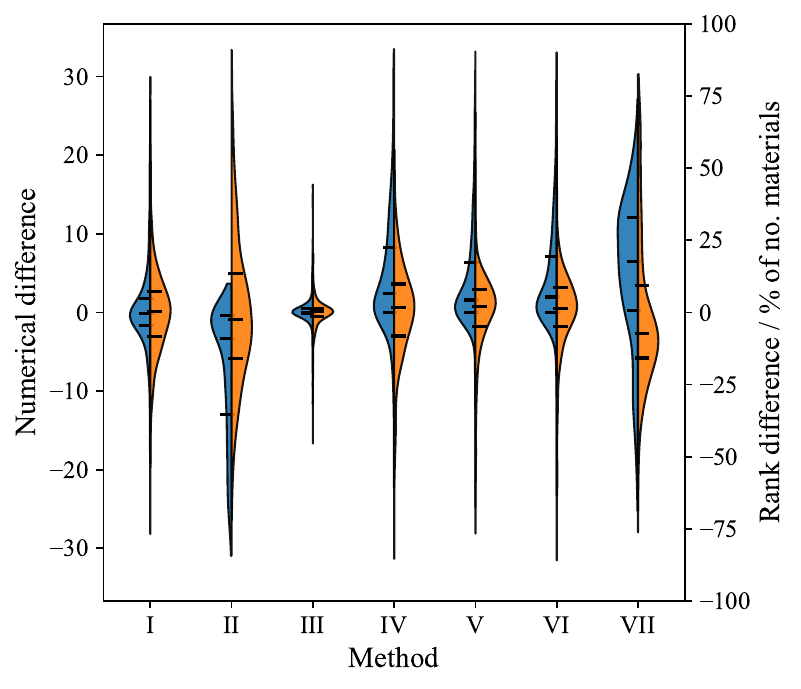}
}
\caption[Fig5]{Violin plot comparing the success of the methods outlined in Table \ref{methods_table} in recreating the test set's SLMEs, in terms of raw accuracy (LH axis) and ranking order when the materials are ranked by their SLME (RH axis). Note that the numerical difference is $\eta_\text{pred}-\eta_\text{true}$ so a positive difference is an overestimate.}
\label{violin_1}
\end{figure}

Figure \ref{violin_1} shows the performance of each method when predicting the SLMEs of the materials in the test set. This performance is quantified by the numerical difference between the target and predicted SLME on the left-hand axis, and the resultant difference in the ranking of these materials when sorted by their SLMEs on the right. This latter distinction is arguably more important when filtering large databases for candidate materials. The difference between these performance quantifications is discussed in more depth in the following section, but we first consider the numerical accuracy.

Comparing the seven methods considered (Method VIII is how the test set is calculated), we see some common trends. Scalar properties (SLME and scissor correction, Methods I and III) are easier to predict than high-dimensional properties (the absorption spectrum as part of Method II). Method II also suffers from the combination of errors, using predictions for the offset (by itself rather well predicted, see Figure \ref{learning_curves}) and the absorption spectrum. This inaccuracy leads the step function-based approaches (Methods IV-VI) to outperform Method II. Otherwise, these approaches struggle compared to direct SLME prediction. Method V, wherein the band gap is calculated at the HSE level, does the best of these approaches, but the cost of this calculation is significantly higher than that of the ML inference in Method I, as discussed in Section \ref{sub_D}. 

Finally, Method VII, based on all GGA-level calculations (without any kind of scissor correction), is the poorest-performing approach. Interestingly, this approach gives the most clearly systematic error, with the vast majority being overestimates. GGA is known to underestimate band gaps due to the self-interaction error, so the absorption profiles will have an earlier offset, and thus we would expect larger short-circuit currents, but not necessarily larger efficiencies due to the voltage-current trade-off: smaller band gaps mean each excited carrier has less energy. We also see some systematic behaviour in Method II, where SLME overestimates are limited to around 5 percentage points, while underestimates can be much more significant. The step function approaches also tend to overestimate SLMEs: this is likely because real absorptance spectra have more gradual onsets than step functions, especially for materials with indirect band gaps.

\subsection{Better accuracy doesn't always give better screening}

The ultimate goal in terms of PV materials discovery might be an accurate direct estimate of SLME from material structure and our previous analysis makes it clear that there is still significant room for improvement. We now consider how the errors in prediction accuracy relate to ranking errors and how this changes for different ML interventions.

We can see from Figure~\ref{violin_1} that different ML interventions introduce errors with different degrees of systematicity. This is a reminder that training objectives and benchmarks commonly used to compare ML models are not always appropriate for a given task \cite{alampara_lessons_2025}. More specifically related to ML for PV screening, this shows that trying to learn SLME directly is probably preferable to prediction of an absorption profile and using that to calculate the SLME. The direct SLME prediction is both more likely to improve with more data and gives more systematic errors. Any effort to generate more high-quality absorption profiles could be trivially translated to SLMEs; therefore, this is the most promising path for the screening of PV materials.

\subsection{There will always be a cost-accuracy trade-off, but there are some sweet spots}
\label{sub_D}
\begin{figure*}[htbp!]
\centering
\scalebox{1.3}[1.3]{
\includegraphics[width=0.75\textwidth]{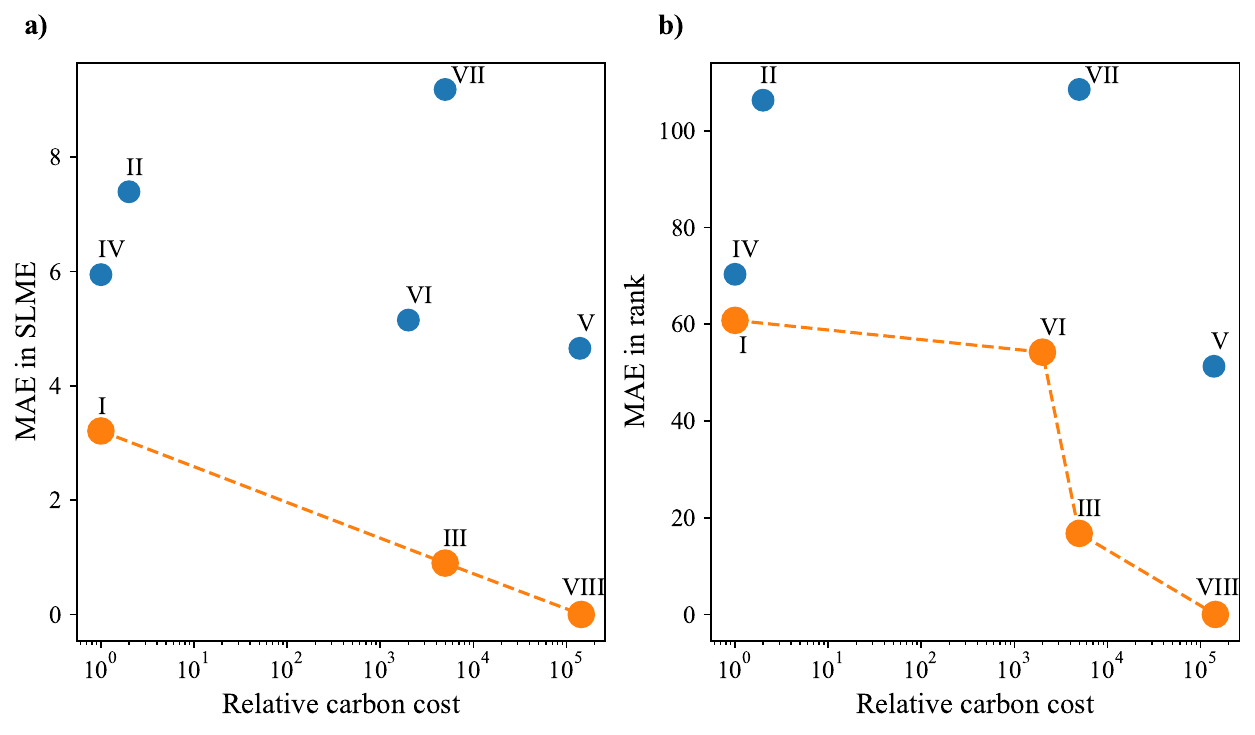}
}
\caption[Fig5]{Pareto front for performance vs cost for the methods outlined in Table \ref{methods_table}, where performance is measured as a) MAE in SLME and b) MAE in rank when the test set and the predictions are sorted by their SLMEs.}
\label{fig:pareto_fronts}
\end{figure*}
Figure \ref{fig:pareto_fronts} (a) shows the 8 methods compared by their cost and accuracy, where the latter is measured as MAE from the held-out test set, so an ideal technique would be close to the origin. Three approaches (I, III, and VIII) stand out as significantly better than the rest, forming a roughly straight line when the cost is logarithmically scaled. Unsurprisingly, the machine learning approaches that predict scalar quantities are similar in cost but much more accurate than the approaches predicting high-dimensional spectra. However, the plot demonstrates the unsuitability of step-function approaches when there exist ML models that can predict SLMEs cheaply and more accurately. Method III, using a learned scissor correction alongside GGA-level absorption spectra and offsets, emerges as a viable intermediate method, nearly two orders of magnitude cheaper than fully hybrid calculations with MAEs under 1.0  percentage points. It should be emphasised that these errors are relative to the fully hybrid approach, which is itself limited. In the SI, we provide a plot of the Method I model applied to a set of 29 high-efficiency materials, whose SLMEs were calculated using GW routines by Yu and Zunger~\cite{yu_identification_2012} and were not in the training set; emphasising the need for high fidelity and large volumes of training data. 

Another consideration when comparing the accuracy of different machine learning approaches is interpretability: the direct SLME prediction is rather a black box, where predicting the absorption spectrum and offset gives us better insight into why a given material is a good absorber. It also allows us to calculate properties like the short-circuit current and photovoltage of a material, extending the applicability of this approach beyond traditional solar cells. Moreover, calculated SLMEs have the temperature, material thickness, and incoming radiation profile (typically the AM1.5G spectrum) implicit in their value, whereas predicting the spectra allows the user to alter these parameters for their application. This could be particularly useful when looking for materials for solar cells used on satellites or in indoor lighting. However, the distance of the Pareto front from the other 5 methods makes it hard to justify this approach. Method III is perhaps the best compromise between interpretability and accuracy.

A final additional factor that could be considered is domain expertise. For instance, comparing V and VIII, we see that if a hybrid band gap is being calculated, it is only slightly more expensive to calculate the GGA absorption spectrum and offset to facilitate an SLME rather than just detailed balance calculation. However, it requires the user to have experience with optics calculations in, for instance, DFT. Packages like \verb|atomate2|~\cite{ganose_atomate2_2025}, used for some example calculations in this report, make this very straightforward, while ML models like the Atomistic Line Graph Neural Network (\verb|ALIGNN|)~\cite{choudhary_atomistic_2021} used in this work are increasingly easy to use out-of-the-box.

Figure \ref{fig:pareto_fronts} (b) tells a similar story, although the difference in numerical accuracy and ranking accuracy is highlighted by Method VI becoming part of the Pareto front. This seems to be a combination of Method I being relatively poor at accurate ranking and Method VI relatively good. However, VI is only narrowly better than I and is over 3 orders of magnitude more expensive, while III is much more accurate at less than 10x the expense, making VI difficult to justify in most instances.

\subsection{Machine learning isn't perfect - but neither is DFT}
\begin{figure}[htbp!]
\centering
\scalebox{0.5}[0.5]{
\includegraphics[width=0.9\textwidth]{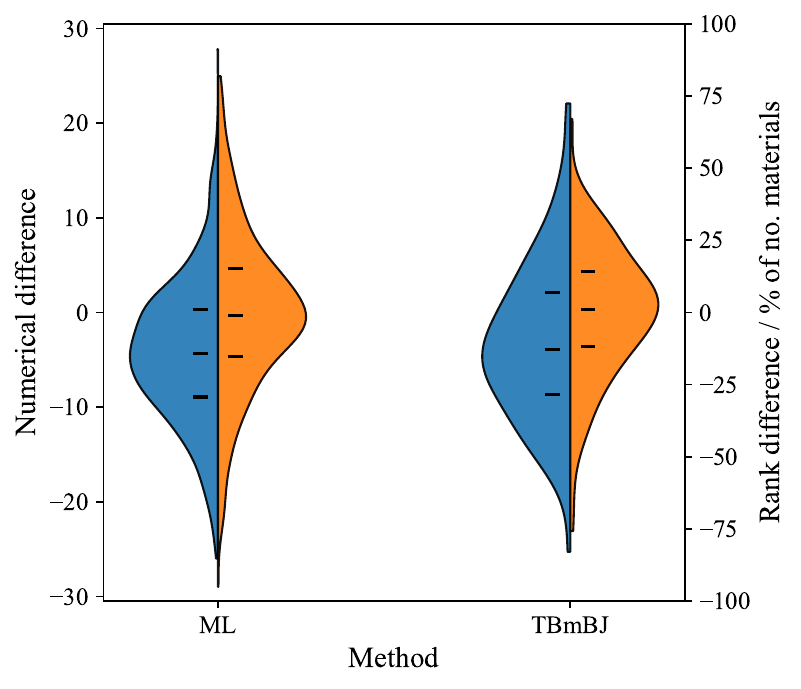}
}
\caption[Fig5]{Violin plot comparing the successes of the SLME-predicting ML model (Method I) and the Choudhary et al.~\cite{choudhary_computational_2018} TB-mBJ dataset in reproducing the SLMEs of an external test set: the Fabini \(\Delta\)-sol set.} 
\label{fig:functionals_violin}
\end{figure}
Next, we evaluate the performance of the direct SLME prediction model (Method I) on an external test set drawn from the work of Fabini et al.~~\cite{fabini_candidate_2019}, which applies the $\Delta$-sol correction scheme of Chan and Ceder~~\cite{chan_efficient_2010} to GGA-level calculations. This test set is entirely independent from the training data used in this work, as it originates from a different set of DFT calculations and computational parameters. As such, it provides a robust test of how well our ML model generalises beyond the specific data distribution upon which it was trained.

Unsurprisingly, the model’s performance on this external dataset is somewhat worse than on the internal test set sampled from the same DFT workflow as the training data. This degradation is expected, as discrepancies between the DFT methodologies used to generate training and test labels introduce additional sources of error, which compound with those from the ML model itself. Nevertheless, the model maintains a reasonable ability to rank materials by predicted SLME, as shown in the rank correlation plots (Figure~\ref{fig:pareto_fronts}).

To contextualise these errors, we also compared SLME values for the same materials computed using two different DFT approaches: $\Delta$-sol-corrected GGA (from Fabini et al.~\cite{fabini_candidate_2019}) and the Tran-Blaha modified Becke-Johnson (TB-mBJ) potential~\cite{tran_accurate_2009} (dataset from Choudhary et al.~~\cite{choudhary_computational_2018}). Interestingly, the absolute and ranking errors between these two DFT methods are comparable in magnitude to the errors observed between the ML predictions and the $\Delta$-sol data. For example, the mean absolute error (MAE) in SLME values between TB-mBJ and $\Delta$-sol is 7.2 percentage points, versus 6.8 percentage points for the ML predictions; similarly, ranking errors are also of similar scale.

These results highlight two important conclusions. First, they demonstrate that the predictive performance of ML models trained on high-fidelity data can approach the level of variability introduced by changes in the DFT methodology itself. Second, they emphasise that the generation of consistent, high-fidelity SLME datasets remains a major bottleneck in data-driven PV discovery. For SLME prediction tasks, our findings suggest that investing in better-quality training data may yield greater improvements than simply expanding the size of existing datasets. In contrast, for absorption spectrum prediction—where model errors remain large even on consistent data—improvements in model architecture and training volume may be the more effective path forward.

\subsection{Ways forward: more data, better data, or better models?}
We now consider how we could push the boundary of efficient PV materials discovery.
One promising avenue for improving ML predictions is through the development of models that more effectively capture the underlying structure–property relationships in the data. In the field of machine-learned interatomic potentials, for instance, the inclusion of physically motivated inductive biases—such as equivariance—has enabled models to achieve high accuracy with relatively modest training datasets~\cite{geiger_e3nn_2022,geiger_euclidean_2022,thomas_tensor_2018,weiler_3d_2018,kondor_clebsch-gordan_2018,batatia_design_2025,batzner_e3-equivariant_2022,musaelian_learning_2023}. Meanwhile, recent attempts~\cite{ruff_connectivity_2024} to optimise the connectivity in the chemical graph used in a GNN have been shown to improve performance relative to the \verb|ALIGNN| model used herein. Similar ideas are beginning to be explored in the prediction of optical properties. 

While these recent efforts show encouraging progress, there are still important limitations. For example, two recent studies~\cite{grunert_deep_2024,hung_universal_2024} have proposed neural network (NN) models for predicting absorption spectra, both demonstrating reasonable accuracy. However, these models were trained and tested on more constrained datasets than the one used in this work, and their performance may degrade when applied to more chemically diverse materials such as those in the W-R dataset. Grunert et al.~\cite{grunert_deep_2024} limited their materials to main-group elements from the first five rows of the periodic table, while Hung et al.~\cite{hung_universal_2024} allowed a broader range of elements but restricted their dataset to structures with nine atoms or fewer per unit cell. Such constraints significantly reduce the overlap with the datasets used here, particularly where both band gap and spectrum data are needed. When trained on the dataset used in this work, the \verb|GNNOpt| model from Hung et al., based on the equivariant \verb|e3nn|\cite{geiger_e3nn_2022,geiger_euclidean_2022,thomas_tensor_2018,weiler_3d_2018,kondor_clebsch-gordan_2018}, predicts spectra that give better SLMEs than \verb|ALIGNN| (see SI), but not enough to become a viable strategy, especially when the errors are confounded with those of the offsets. This suggests that developments in model architectures such as these will continue to drive improved predictions of PV-relevant properties.

Another key challenge lies in the availability of consistent, high-quality training data. Both of the recently proposed neural network models for spectral prediction were trained on data generated using generalized gradient approximation (GGA) functionals, which—as we have shown—can lead to suboptimal screening performance. The reliance on GGA is largely driven by its relative abundance compared to more accurate methods, such as hybrid-DFT. However, progress in data infrastructure and learning techniques offers promising ways forward. Initiatives such as the Novel Materials Discovery (NOMAD) program~\cite{sbailo_nomad_2022} and MPContribs (the platform for contributing to the Materials Project~\cite{jain_commentary_2013}) are enabling the sharing of curated, high-quality computational datasets in line with FAIR data principles~\cite{wilkinson_fair_2016}. 

At the same time, recent advances in multi-fidelity machine learning~\cite{fare_multi-fidelity_2022,hoffmann_transfer_2023,kaur_data-efficient_2025} allow models to be trained on datasets that combine varying levels of theoretical accuracy. By leveraging correlations between low- and high-fidelity data, these methods enable the use of larger training sets without sacrificing predictive reliability, thereby offering a practical route to more robust and generalizable ML models for materials discovery.

Closely tied to the development of better data and models is the need for high-quality community benchmarks. As our results demonstrate, benchmarking efforts should not only assess predictive performance but also account for the environmental cost of computation—such as carbon emissions—which can meaningfully influence the practicality of different approaches. Evaluation choices fundamentally shape not only our measurements, but also research priorities and scientific progress. Ensuring transparency and reproducibility in benchmarking is therefore critical. Recent proposals, such as evaluation cards, offer a structured means of documenting the assumptions, metrics, and limitations that underpin model assessments~\cite{alampara_lessons_2025, mitchell_model_2019}. By adopting such practices in the context of materials discovery, the community can move toward more robust, equitable, and environmentally conscious progress in the development of machine learning for photovoltaics and beyond.

A final consideration for improvement is the SLME metric itself. The Blank selection metric~\cite{blank_selection_2017} has emerged as a more accurate computational characterisation of photovoltaic efficiency. However, it requires additional data such as the refractive index $n(E)$, of which there are currently no large datasets. A more rigorous computational study of a candidate photovoltaic would go even further, considering factors such as defects, dopants, and stability under real operating conditions. However, as a heuristic for filtering large areas of chemical space for intrinsically good PV absorbers, the SLME should be sufficient, hence its use in this work. As we have emphasised with the ranking plots, exact numbers for efficiency are less important than identifying the best materials.
\section{Conclusion}
ML has the potential to dramatically accelerate the discovery of new PV materials. However, as we demonstrate here, it is not (currently at least) a panacea. Current limitations mean that for successful materials discovery campaigns for thin-film PV a combination of ML surrogates and electronic structure calculations is required.  Our findings suggest that direct prediction of the SLME offers the most cost-effective approach to obtain reliable estimates of photovoltaic performance. Similarly, learned scissor corrections can substantially improve the accuracy of GGA absorption spectra at a fraction of the computational cost required for HSE band gap or optics calculations. However, direct spectral prediction currently introduces too much error to be practically useful for discovering novel photovoltaic materials, despite the appeal of the flexibility of this approach.

We have also identified clear pathways to improve ML surrogate models. Enhanced performance will likely require either substantially larger datasets of high-fidelity calculations than are presently available, or the implementation of transfer learning approaches that leverage extensive low-fidelity datasets alongside smaller, high-accuracy training sets.

More broadly, our study highlights the fundamental trade-off between computational cost and the efficacy of data-driven screening in materials design. We have outlined a blueprint for jointly evaluating the carbon cost and discovery performance of such campaigns. Embedding carbon cost reporting into computational discovery workflows is, we argue, a vital step toward ensuring that AI-, ML-, and simulation-driven approaches deliver truly beneficial and socially responsible innovation.

\section{Methods}
The Atomistic Line Graph Neural Network (\verb|ALIGNN|) model~\cite{choudhary_atomistic_2021} was used for ML predictions, with output dimensions of 1 or 100 for the scalar and spectral properties, respectively. Spectral data were represented by binning into 100 dimensions using the \verb|numpy.interpolate()| function rather than compression into latent dimensions using, for example, a variational autoencoder~\cite{kingma_auto-encoding_2022}. This decision was largely based on a paper from Kaundinya et al.~\cite{kaundinya_prediction_2022} that used \verb|ALIGNN| to predict the electron density of states of inorganic materials and found the binning approach (into 300 bins in their case) to be the more successful of the two.  

Z-score normalisation was used to scale labels for a more stable gradient descent; spectral properties were normalised per bin. Each model was trained for 300 epochs with a batch size of 64 and the rest of the hyperparameters in line with the model's original paper for consistency across the various properties predicted. A batch size of 2 was used for the learning curves as this enabled each dataset size to be trained with the same batch size. 

Datasets from Woods-Robinson et al.~\cite{woods-robinson_designing_2023}, Kim et al.~\cite{kim_band-gap_2020}, Fabini et al.~\cite{fabini_candidate_2019}, Yu and Zunger~\cite{yu_identification_2012}, and Choudhary et al.~\cite{choudhary_computational_2018} were used, all accessed from freely available sources. The main dataset (the $\sim$5.3k overlapping materials from W-R and Kim) was split into an 80:10:10 ratio of training:validation:test data; the test materials were kept the same for all models for fairer comparison. 

Some examples of DFT calculations at GGA and HSE levels were performed using the projected augmented wave (PAW) method~\cite{kresse_norm-conserving_1994,kresse_ultrasoft_1999} within the Vienna \textit{ab initio} Simulation Package (VASP)~\cite{kresse_ab_1993,kresse_efficiency_1996,kresse_efficient_1996}, with \verb|CodeCarbon|~\cite{courty_mlco2codecarbon_2024} monitoring the energy (and thus carbon) cost of each calculation. \verb|Atomate2|~\cite{ganose_atomate2_2025} was used to generate the input files for these calculations, with structure files from the Materials Project~\cite{jain_commentary_2013}, to simulate a high-throughput workflow rather than bespoke calculations for each material.  The raw numbers for these costs are available in the SI. \verb|CodeCarbon| was also used for some ML training and inferences. \verb|Matplotlib| was used for plotting. 
\section{Code and data availability}
The code and data used in this work are provided at \url{https://github.com/mattheww98/PV_paper.git}.
\section{Acknowledgements}
We acknowledge support from EPSRC project EP/Y000552/1 and  EP/Y014405/1. Via our membership of the UK's HEC Materials Chemistry Consortium, which is funded by EPSRC (EP/X035859/1), this work used the ARCHER2 UK National Supercomputing Service \url{http://www.archer2.ac.uk}. The authors acknowledge the use of the UCL Myriad High Performance Computing Facility (Myriad@UCL), and associated support services, in the completion of this work. We acknowledge Young HPC access which is partially funded by EPSRC (EP/T022213/1, EP/W032260/1 and EP/P020194/1).  We acknowledge the assistance of the MPcontribs team, in particular Patrick Huck for assisting us in assembling our initial datasets.

\bibliography{bibliography}

\end{document}